\PassOptionsToPackage{square,comma,numbers,sort&compress,super}{natbib}
\documentclass[preprint,12pt,3p]{elsarticle}
\usepackage[utf8]{inputenc} 
\usepackage[T1]{fontenc}    
\usepackage{hyperref}       
\usepackage{url}            
\usepackage{booktabs}       
\usepackage{amsfonts}       
\usepackage{nicefrac}       
\usepackage{microtype}      
\usepackage{lipsum}
\usepackage{amsmath,amssymb,graphicx}
\usepackage{siunitx}
\usepackage[utf8]{inputenc}
\usepackage[T1]{fontenc}
\usepackage{array}
\usepackage{lmodern}

\journal{Elsevier}
\begin{document}
\begin{frontmatter}
\title{Boron Nitride Nanotube Peapod under Ultrasonic Velocity Impacts: A Fully Atomistic Molecular Dynamics Investigation}
\author[IFPI,UNICAMP]{J. M. De Sousa\corref{author}}
\ead{josemoreiradesousa@ifpi.edu.br}
\cortext[author]{I am corresponding author}
\address[IFPI]{Instituto Federal de Educação, Ciência e Tecnologia do Piau\'i -- IFPI, Primavera,
 S\~ao Raimundo Nonato, Piau\'i, 64770-000, Brazil.}
 \author[UFRN]{L. D. Machado }
\address[UFRN]{Departamento de Física Teórica e Experimental, 
  Universidade Federal do Rio Grande do Norte -- UFRN, 
  Natal, 59072-970, RN, Brazil.}
\author[UFPR]{C. F. Woellner }
\address[UFPR]{Physics Department, Federal University of Parana – UFPR, Curitiba, 81531-980, PR, Brazil.}
\author[UFSP-ABC]{M. Medina }
\address[UFSP-ABC]{Center of Natural Human Science, 
  Federal University of ABC -- UFABC,
  Santo Andre, 09210-580, SP, Brazil.}
\author[UFSP-ABC]{P. A. S. Autreto }
\author[UNICAMP]{D. S. Galvão}
\address[UNICAMP]{Applied Physics Department, State University of Campinas -- UNICAMP,
 Campinas, 13083-859, SP, Brazil.}  

\begin{abstract}

In this work, we investigated the mechanical response and fracture dynamics of boron nitride nanotubes (BNNTs)-peapods under ultrasonic velocity impacts (from $1$ km/s to $6$ km/s) against a solid target. BNNT-peapods are BNNTs containing an encapsulated linear arrangement of C$_{60}$ molecules.  We carried out fully atomistic reactive (ReaxFF) molecular dynamics simulations. We have considered the case of horizontal and vertical shootings. Depending on the velocity values we observed tube bending, tube fracture, and C$_{60}$ ejection. One interesting result was tube unzipping with the formation of bilayer nanoribbons 'incrusted' with C$_{60}$ molecules.


\end{abstract}


\end{frontmatter}
\section{Introduction}
\label{INT}

The hypervelocity impacts (HVI) of single- and double-walled carbon and boron nitride nanotubes have been extensively studied, theoretically and experimentally, in the past decade. In these experiments, the nanotubes collide with solid substrates at ultrasonic velocities (usually between $2$ and $8$ km/s). Impact forces induce significant deformations that can result in a significant number of new structures, depending on many variables, such as impact angle and velocity, number of layers, material composition, among others \cite{li2008chemically,jiao2009narrow,kosynkin2009longitudinal,ozden2014unzipping,ozden2016ballistic,machado2016structural}.

For single-walled carbon nanotubes (SWCNT), the high-velocity impacts generate (due to an internal folding) localized stress on the ends of the tubes, breaking C-C bonds in an almost perfect line, unzipping the SWCNT into carbon nanoribbons \cite{ozden2014unzipping}. Similarly, double-walled carbon nanotubes (DWCNT) become completely unzipped (two walls at the same time) through stress accumulation at the edges of the structure, as internal folding is constrained by the second wall \cite{dos2012unzipping}. On the other hand, double-walled boron-nitride nanotubes (DWBNT) unzip one wall at a time, due to the higher hardness of the inner core \cite{perim2013dynamical}. 

All previously cited studies were carried out for nanotubes in which the external and internal walls have the same compositions (carbon or boron nitride). Experimentally, hybrid tubes with different wall compositions (external and internal) have been already synthesized. One example is the hybrid BN-C nanotube (composed of concentric carbon and boron-nitride walls), experimentally realized by Nakanishi’s group \cite{nakanishi2013thin}. A study of highly energetic impacts of these new hybrid nanotubes has been already reported, where the authors predicted the elastic response of the individual walls \cite{armani2021high}. The authors showed that they have remarkably different elastic properties in relation to CNT and BNNT. For example, the tubes can decouple, and the carbon nanotube can unzip. These results open up a new perspective for the synthesis of nanostructures with different topologies and compositions.

An interesting example of a hybrid carbon-based nanostructures are the carbon nanotube peapods, which are composed of fullerenes encapsulated into carbon nanotubes. They have been synthesized through different experimental techniques \cite{smith1998encapsulated,burteaux1999abundance,bandow2001raman,smith2000formation}, and exhibit very interesting chemical and physical properties \cite{muramatsu2009bright,pfeiffer2004electronic,rochefort2003electronic}, and even higher thermal conductivity than SWCNT \cite{noya2004thermal}. These remarkable properties have been exploited in many applications in nanotechnology, which include transistors\cite{shimada2002ambipolar}, solar cells\cite{hatakeyama2010infrared}, nano-memory devices\cite{lee2003nano}, and supercapacitors\cite{jiang2011peapod}. A previous HVI study \cite{de2020carbon} showed that peapods exhibit remarkable resilience under high strain rate conditions. The calculations show large structural deformation and multiple fracture pathways, depending on the impact velocity and relative nanotube-substrate orientation during the impact.

\begin{figure}[ht!]
\begin{center}
\includegraphics[angle=0,scale=0.50]{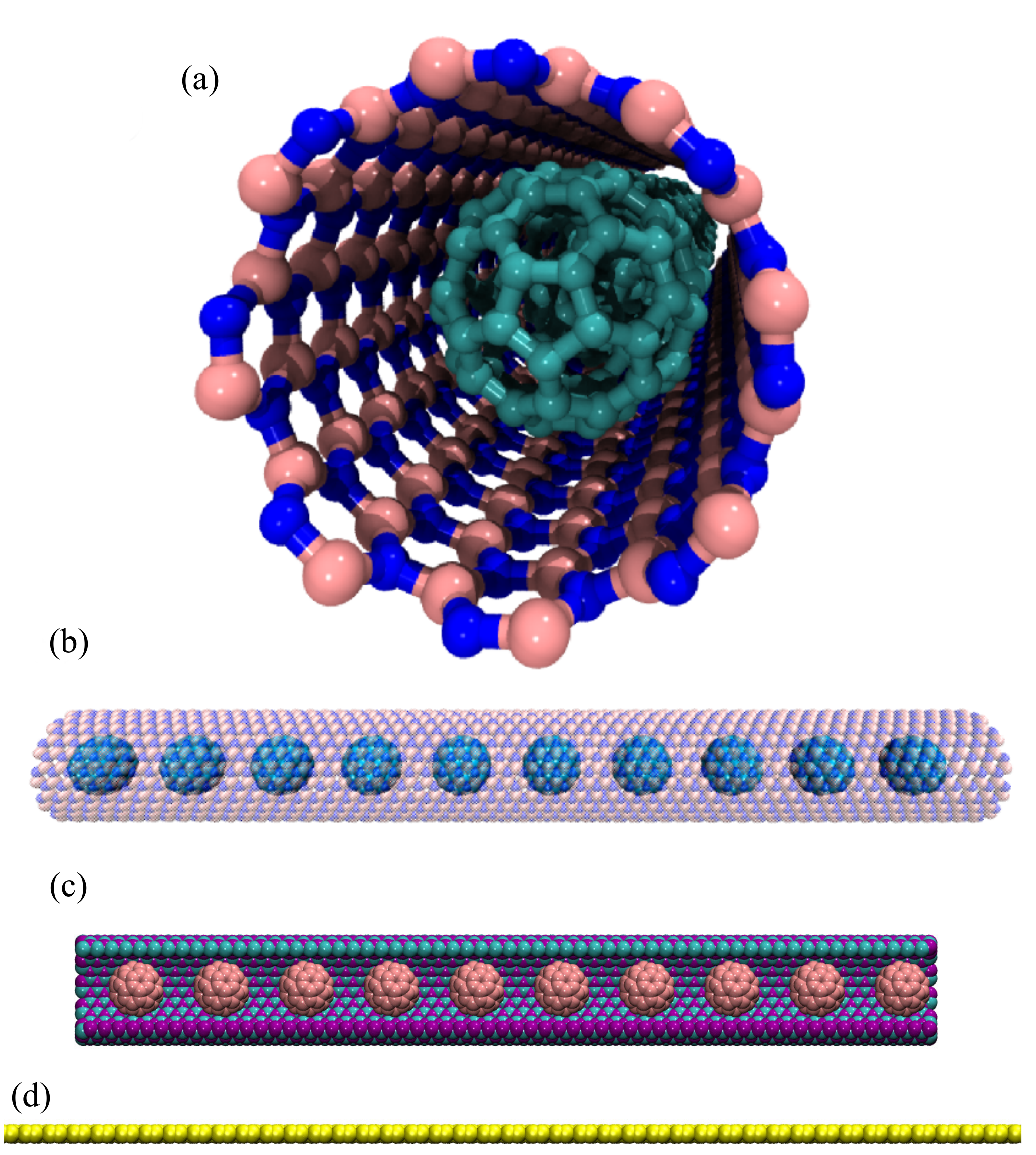}
\caption{\footnotesize{Representative structural atomic model of a boron nitride peapod (BNNT-peapod). (a) Frontal view, and (b-c) side views of a BNNT-peapod with ten encapsulated C$_{60}$ fullerenes. In (c), part of the outer wall was made transparent for a better view. (d) Side view of the impact target. See text for discussions.}}
\label{fig01}
\end{center}
\end{figure}

In this work, we investigated another hybrid structure, BNNT-peapods, which are nanostructures composed of C$_{60}$ fullerenes encapsulated into boron nitride nanotubes  \cite{kitaura2007endohedral}(Fig. \ref{fig01} (a-c)). These structures have been already synthesized by Mickelson et al. \cite{mickelson2003packing}. 

\vspace{5em}

\section{Computational Methodology}
\label{CM}

 
\begin{figure}[htpb]
\begin{center}
\includegraphics[angle=0,scale=0.45]{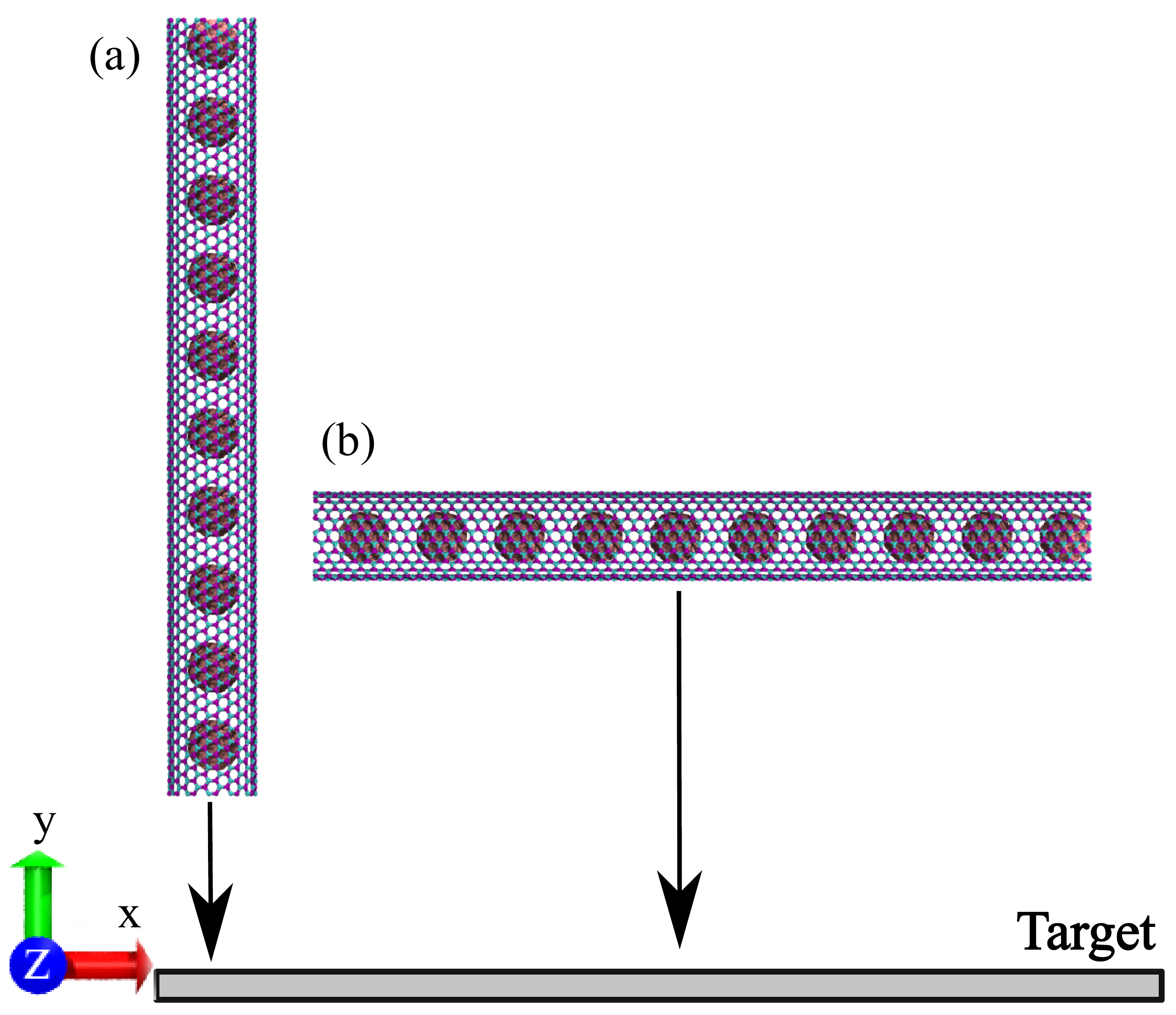}
\caption{\footnotesize{Representative atomic models, showing the considered orientations between the BNNT-peapods and the target substrate. (a)Vertical axis orientation ($90^{\circ}$ with the target) and (b) lateral axis orientation ($0^{\circ}$ with the target). }}
\label{fig02}
\end{center}
\end{figure}

In this work, we carried out molecular dynamics (MD) simulations to analyze the fracture patterns of BNNT peapods. We have considered two impact orientations (see Figure \ref{fig02}(a-b)): (a) vertical shooting, where the angle between the BNNT main axis and the plane of the target is $90^{\circ}$; (b) lateral shooting, where the angle is $0^{\circ}$.  For both cases, we considered impact velocity starting from $1$ km/s (low speed) up to $6$ km/s (ultrasonic speed), with a step velocity increment of $1$ km/s. Fully atomistic reactive molecular dynamics simulations were used to characterize the mechanical response of BNNT peapods, their fracture dynamics, and structural deformations under impacts against a rigid target.


The simulations were carried out using the set of parameters for $H/C/B/N$ atoms defined for the Reactive Force Field (ReaxFF), as implemented in the LAMMPS code \cite{plimpton1995fast}. The reactive force field allows the breaking and formation of chemical bonds, and it was previously used to study the mechanical behavior of nanostructures under ultrasonic impacts \cite{paupitz2014fullerenes,de2016carbon,woellner2018structural}. ReaxFF produces results in good agreement with quantum simulation methods in describing carbon-based nanostructures \cite{nielson2005development,van2001reaxff}.

For all simulations, we considered a $14.9$ nm long $(12,12)$ BNNT, containing an encapsulated linear arrangement of $10\: C_{60}$ molecules. In this configuration, the distance among the C$_{60}$ is $0.837$ nm (see Figs. \ref{fig02} (a) and (b)). In total, the hybrid system contains $3480$ atoms. The target used in the simulations is a rigid van der Waals wall. A time step of $0.025$ fs was used to integrate the equations of motion. The BNNT-peapods were first equilibrated in an NVT ensemble at $300$ K for $1000$ time-steps, using a chain of three Nosé-Hoover thermostats \cite{evans1985nose}. Then, we turn off the thermostat, assign a vertical velocity to the system, and let it freely evolve for $5\times10^5$ time steps in the NVE ensemble. This approach has been successfully applied to many theoretical investigations of the hyper-velocity impact of nanostructures\cite{de2016carbon,woellner2018structural,ozden2014unzipping,armani2021high}.

We analyzed in details the BNNT-Peapods during and after the impacts. During the impact, the BNNT-peapods experience pressure-induced deformations, the pressure values can be evaluated, as a function of time, using the following expression: \cite{buehler2008atomistic,landau1986theory}:
\begin{eqnarray}
P_{atoms} =  Nk_{B}T - \left( \frac{1}{V} \right)\frac{1}{3}\left[ \sum_{l=1}^{N} \sum_{m=1,m>l}^{N} \left \langle r_{lm}\frac{d\phi}{dr_{lm}} \right \rangle\right].
\label{press}
\end{eqnarray}
Where in the first term $N$ is the number of atoms, $k_b$ is the Boltzmann constant, and $T$ is the temperature. The second term is related to the interatomic forces among the atoms, whose elements are the contribution of the force and the position vector components between the $l$ and $m$ atoms in the BNNT peapods. We also calculated the time evolution of the hydrostatic stress, which is equal to one-third of the trace of the stress tensor \cite{hill1950mathematical}. The hydrostatic stress can be negative (for compressive stress) or positive (for tensile stress). A similar  methodology was successfully used in other studies \cite{de2016mechanical,de2016torsional,de2019elastic,de2020hydrogenation,de2021computationalPENTA,brandaomechanical,de2021mechanicalDFT}.

\section{Results and Discussions}

\begin{figure}[ht!]
\begin{center}
\includegraphics[angle=0,scale=0.25]{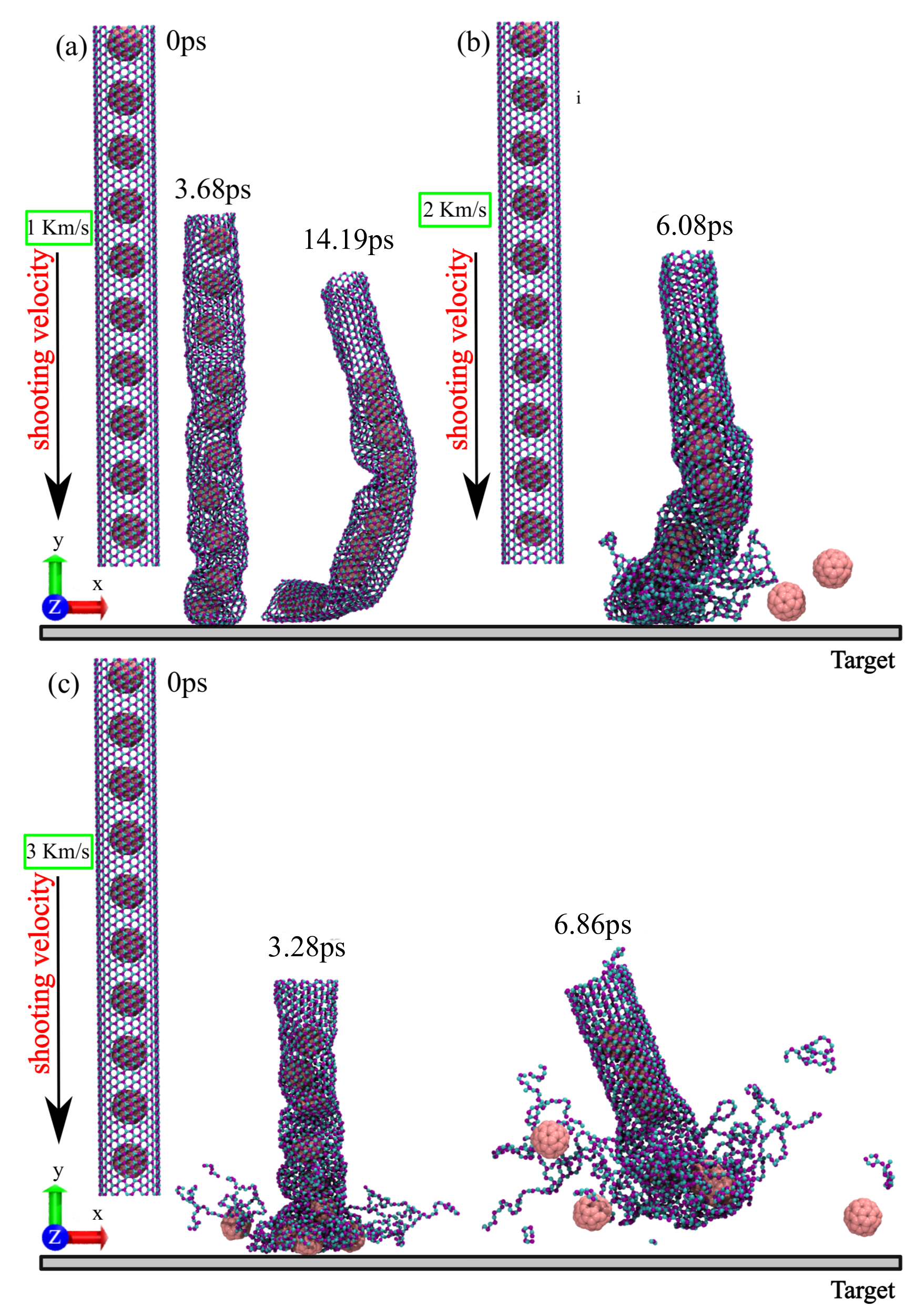}
\caption{\footnotesize{{\textit{Representative MD snapshots for vertical shooting ($90^{\circ}$ degrees with the target) at three different impact velocities values. The numbers presented along each snapshot correspond to the time elapsed since the shooting.}}}}
\label{verticalshooting-snapshots}
\end{center}
\end{figure}

Representative temporal evolution MD snapshots for vertical shooting ($90^{\circ}$ degrees with the target) at $3$ different speeds ($1$, $2$, and $3$ km/s) are presented in Figure \ref{verticalshooting-snapshots}. The results show that under vertical impact the major structural deformations and/or fractures are on the nanotubes, the degree of these deformations/fractures, as expected, depends on the velocity values. For low values ($1$ km/s), the conversion of the kinetic energy into elastic energy primarily causes nanotube bending (Figure \ref{verticalshooting-snapshots} (a)). As the velocity increases, the tube is fractured, with an increasing number of broken bonds (Figure \ref{verticalshooting-snapshots} (b)), resulting in a fast release of the kinetic energy. For v$=2$ km/s, the impact causes both BNNT bending and bond breaking, while for v$=3$ km/s, the impact essentially leads to extensive bond breaking and C$_{60}$ ejection (Figure \ref{verticalshooting-snapshots} (c)). These results are similar to others reported in the literature for tube impact studies showing that the vertical impact can induce extensive fractures due to a high accumulation of energy in a small number of atoms and bonds \cite{de2016mechanical,de2016torsional,de2019elastic,de2020hydrogenation,de2021computationalPENTA,brandaomechanical,de2021mechanicalDFT}.


\begin{figure}[h]
\begin{center}
\includegraphics[angle=0,scale=0.30]{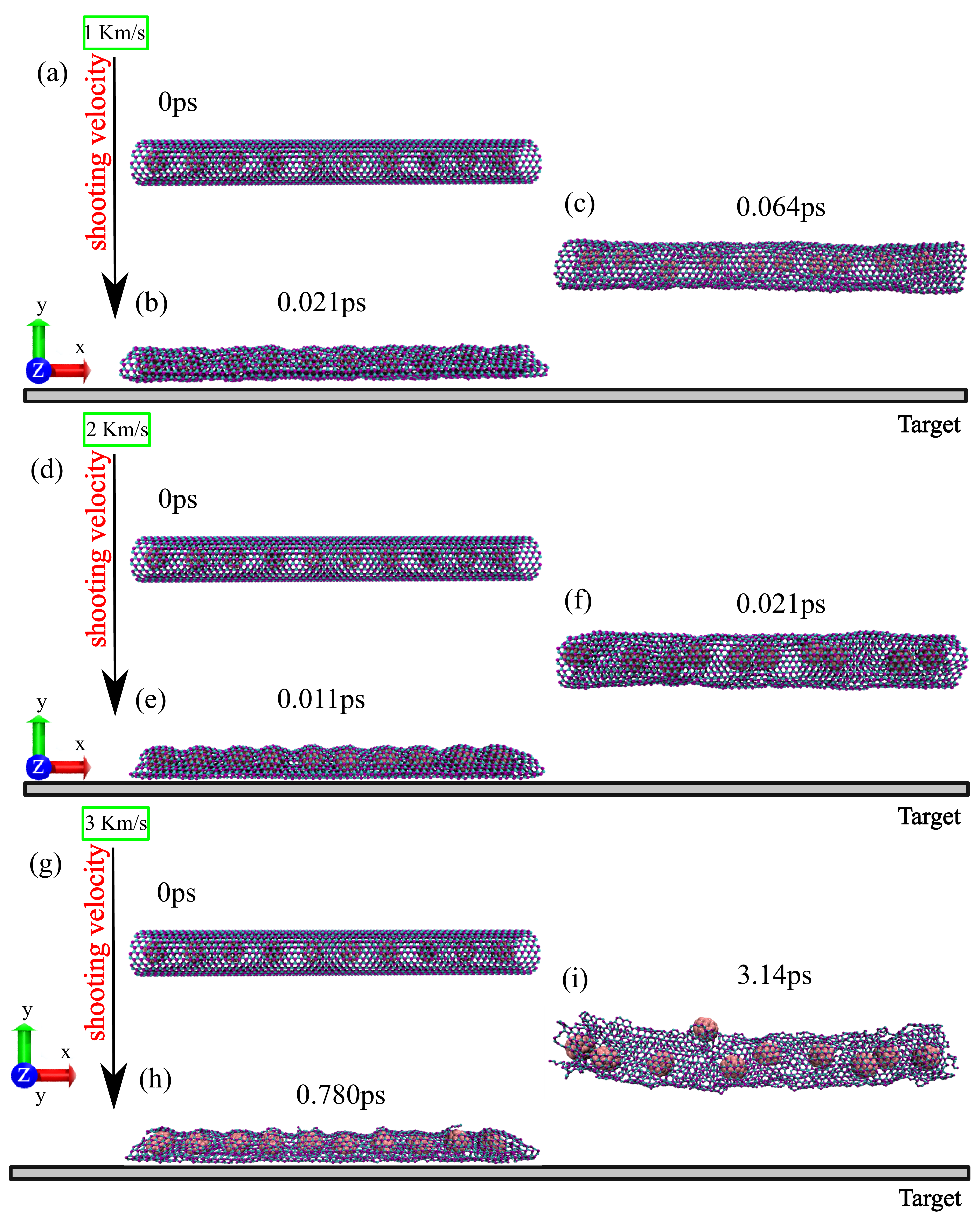}
\caption{\footnotesize{{\textit{Representative MD snapshots for horizontal shooting ($0^{\circ}$ degree with target) at three different impact velocity values. The numbers presented along each snapshot correspond to the time elapsed since the shooting.}}}}
\label{horizontalshooting-snapshots}
\end{center}
\end{figure}

Next, we considered the results for horizontal shooting. In Figure \ref{horizontalshooting-snapshots} are presented some representative MD snapshots for horizontal shooting ($0^{\circ}$ degree with the target). From this figure, it is possible to see that there is a better distribution the kinetic energy during the impact reducing the number of atoms ejected from the structure. The C$_{60}$ remain intact for impact velocities up to 3 km/s. For the BNNTs, the level of structural fragmentation again depends on the impact velocity value. For v$=1$ km/s, the nanotube is deformed without fracture, and for v$=2$ km/s, a few bonds break, leading to the formation of defects on the BNNT wall. On the other hand, for v=$3$ km/s, the BNNT fractures are extensive. Careful inspection of the results reveals that bonds break mainly near the C$_{60}$ inside the nanotubes. We discuss this result in more detail further in the article.


\begin{figure}[ht!]
\begin{center}
\includegraphics[angle=0,scale=0.30]{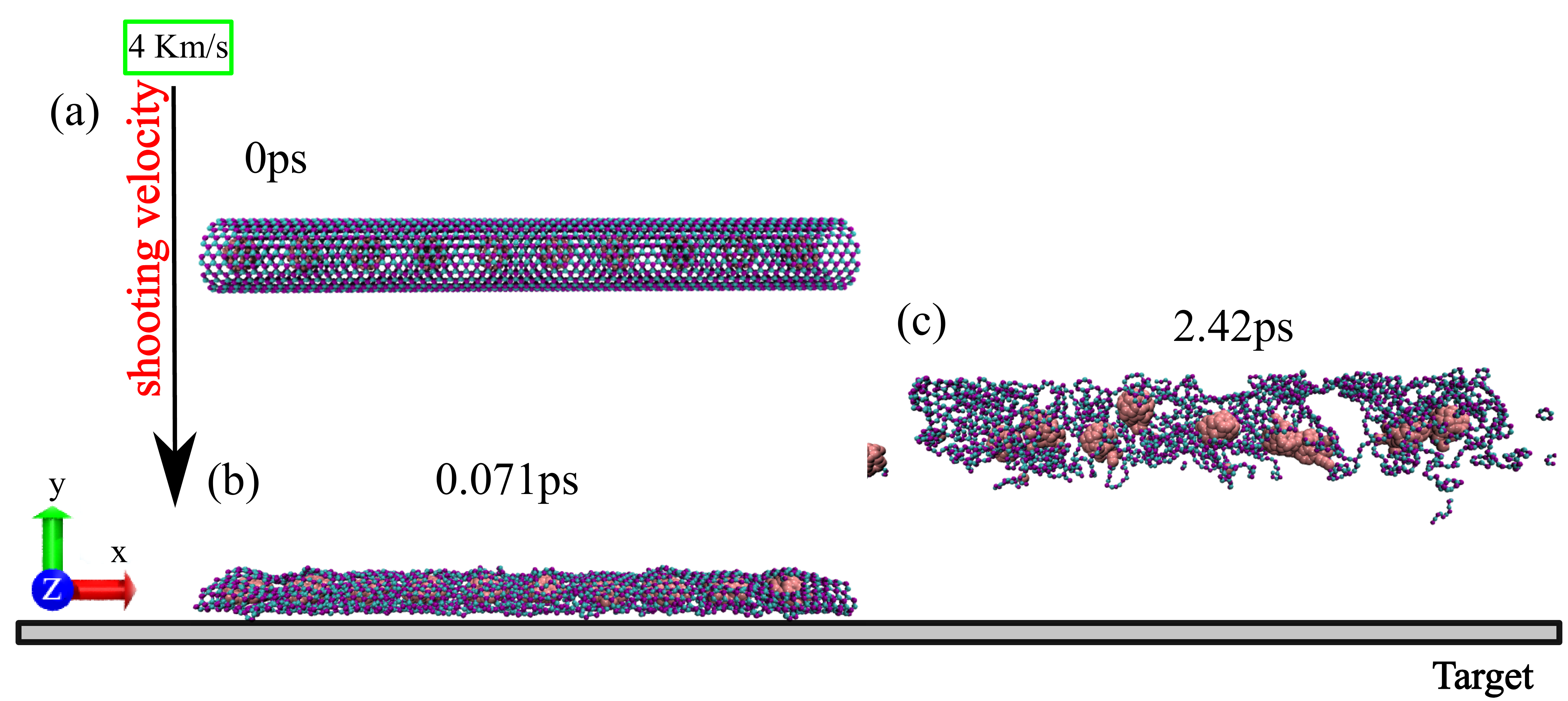}
\caption{\footnotesize{{\textit{Representative snapshots for horizontal shooting ($0^{\circ}$ degree with target) at $4 km/s$. The numbers presented along each snapshot correspond to the time elapsed since the shooting.}}}}
\label{horizontalshooting-4kms}
\end{center}
\end{figure}

\begin{figure}[ht!]
\begin{center}
\includegraphics[angle=0,scale=0.45]{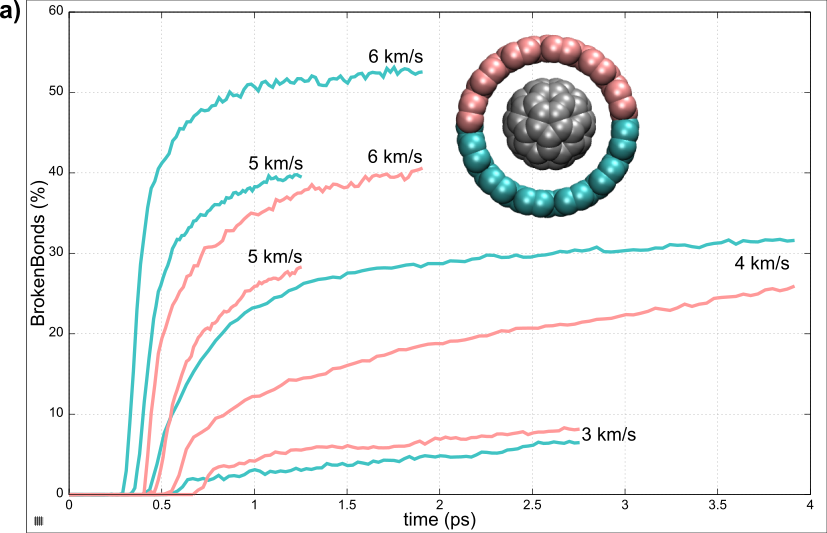}
\includegraphics[angle=0,scale=0.45]{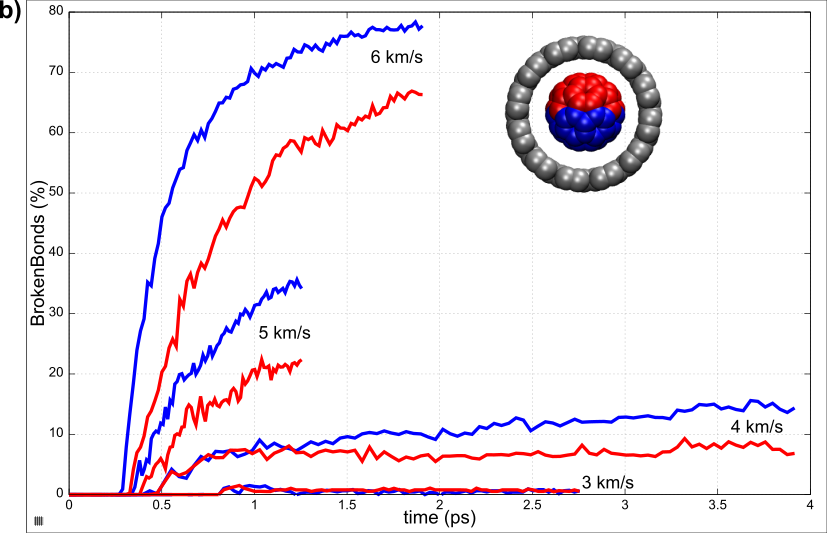}
\caption{\footnotesize{{\textit{Percentage of broken bonds as a function of the simulation time. (a) Results for the BNNT. The color used to display the curve indicates the corresponding region: pink for the top part and cyan for the bottom one (see the inset). The numbers next to each curve indicate the shooting velocity value. (b) Results for the C$_{60}$. The color used to display the curve indicates the corresponding region: red for the top part and blue for the bottom one (see the inset).}}}}
\label{breakingbonds}
\end{center}
\end{figure}

From $4$ km/s (Figure \ref{horizontalshooting-4kms}), the number of broken bonds in the BNNT wall increases, and the impact starts also fracturing the C$_{60}$. In Figure \ref{breakingbonds} we present the percentage of broken bonds as a function of the simulation time, which corroborates this assertion. In this figure, we divided the system into four regions: (i) the bottom and (ii) the top of the BNNT; (iii) the bottom and (iv) the top of the C$_{60}$. These results reveal a clear relationship between impact velocity and fracture patterns. Examining Figure \ref{breakingbonds}(a), for velocity values equal to or below $3$ km/s, there is no predominant region for the fracture of the nanotube. For both, the bottom and top areas, the percentage of broken bonds have a  plateau at around $7\%$. However, as the impact velocity increases, an asymmetry emerges in the bond breakage between the upper and lower regions. For instance, for v $=5$ km/s, the difference is around 11 percentage points. Thus, for higher impact velocities, the presence of C$_{60}$ induces an increase of broken bonds at the bottom of the nanotubes, preserving the top area. This effect enables the possibility of obtaining graphene nanoribbons from the unzipped nanotubes. For the C$_{60}$, a similar asymmetry emerges between the top and bottom regions (Figure \ref{breakingbonds}(b)). As mentioned before, for the most part, the fullerenes remain intact after the impact for velocity values below $4$ km/s. At v $=4$ km/s, around $15\%$ of the bonds at the bottom part of the C$_{60}$ breaks, compared to $32\%$ for the BNNT bottom. On the other hand, for an impact velocity of $6$ km/s, the C$_{60}$ fractures are more extensive than those of the nanotubes. The C$_{60}$ are almost totally fragmented for this critical velocity value.

\begin{figure}[ht!]
\begin{center}
\includegraphics[angle=0,scale=0.11]{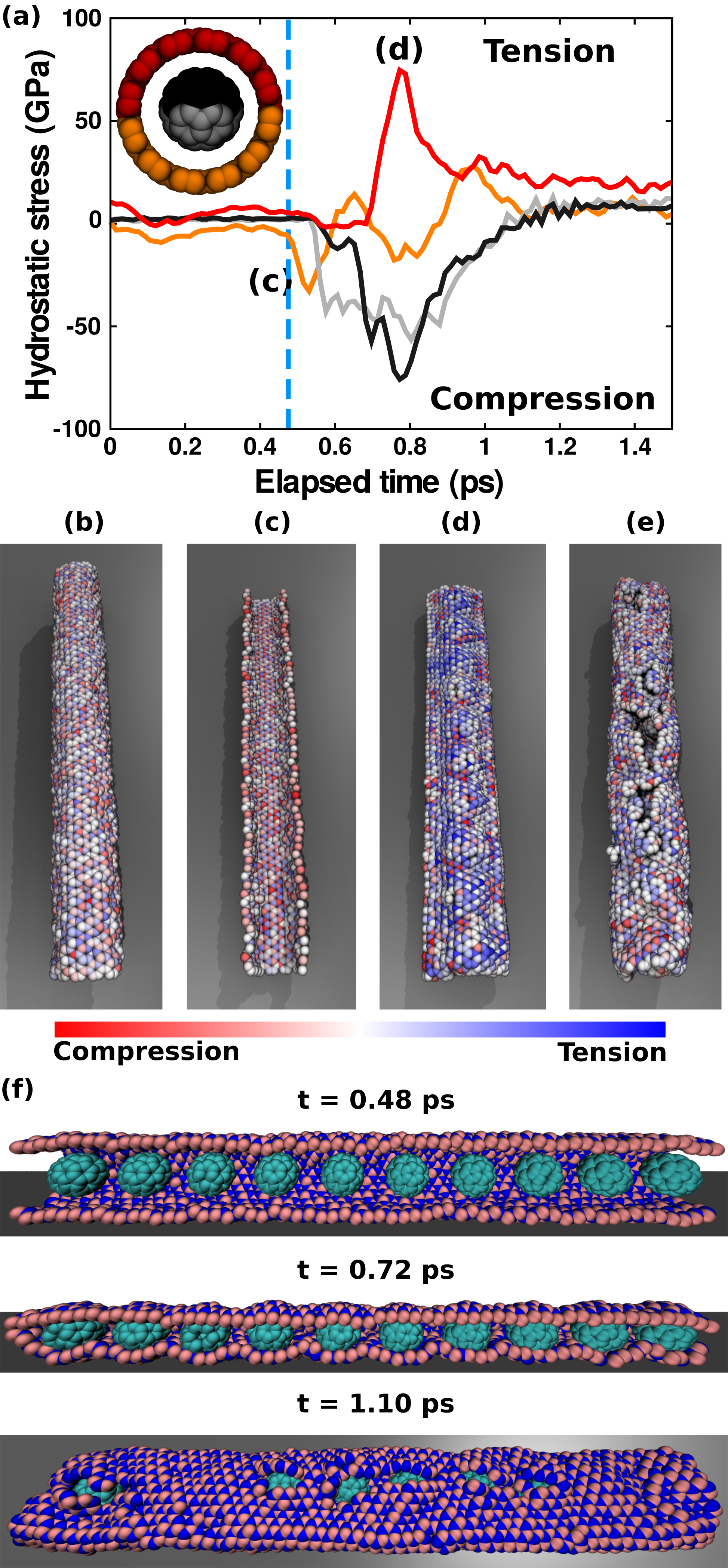}
\caption{\footnotesize{(a) Hydrostatic stress values as a function of the simulation time. Each curve is associated with a corresponding region of the system, illustrated by using the same color in the inset. (b-e) Representative MD snapshots, where the color indicate the local distribution of the hydrostatic stress. (b) and (e) present the initial and final stress configurations, while (c) and (d) present the distribution at the instants highlighted in (a). (f) MD snapshots displaying a side view of the BNNT at three instants, revealing the C$_{60}$ configuration during the impact.}}
\label{fig07}
\end{center}
\end{figure}


BN-peapods at high-velocity impact can also result in unzipped nanotubes. To investigate this possibility, we selected velocity values between $3$ km/s (where the asymmetry in bond breakage is small) and $4$ km/s (where the fracture is already extensive). The results present below are for a shooting velocity of $3.3$ km/s. In Figure \ref{fig07} we present the hydrostatic stress values as a function of the simulated time. We selected an example where the tube unzipping occurred. To investigate the propagation of stress in detail, we divided the system into four regions and calculated the average stress in each part at each instant. In part (a) of Figure \ref{fig07}, the orange/red curves correspond to the average stress in the lower/upper part of the BNNT, while the gray/black curves correspond to the average stress at the lower/upper part of the C$_{60}$. One advantage of using the hydrostatic stress is that its sign differentiates compressive (negative) and tensile (positive) stresses. We also analyzed local stress patterns, and in Figures \ref{fig07}(b-e) we used a color scale ranging from red to blue to indicate the local hydrostatic stress values.   

Before the collision with the substrate (indicated with the blue dotted line in Figure \ref{fig07}(a)), we find that, as expected, the stress values are low throughout the system (Figure \ref{fig07}(b)). Since the lower part of the BNNT collides first with the substrate, immediately after the impact, the stress increases in this region, leading to the first compressive peak in orange in Figure \ref{fig07}(a). Figure \ref{fig07}(c) presents the corresponding stress distribution in this region, and we find that the entire lower part of the BNNT is under compression at this instant. The stress then propagates upward, towards the lower part of the C$_{60}$ (gray peak in Figure \ref{fig07}(a)) and then towards their upper parts (black peak in Figure \ref{fig07}(a)). Directly after this compressive spike, we notice a sharp tensile stress peak in the upper part of the BNNT, reaching 74.6 GPa, as the C$_{60}$ push this region up (see Figure \ref{fig07}(d)). Consequently, a partial fracture occurs through this BNNT region, as shown in Figure \ref{fig07} (e). Figure \ref{fig07}(f) provides further insight on the partial unzipping process. In this Figure, we made part of the nanotube transparent to highlight the C$_{60}$ configuration inside the BNNT during the collision. Notice in the frame obtained at t = 0.72 ps that the C$_{60}$ press against the upper part of the BNNT, leading to fracture at some points of contact at t = 1.10 ps. In all cases, we only observed partial unzipping as a result of the presence of gaps between the instances. Finally, note that the stress values at the BNNT-C$_{60}$ points of contact are certainly higher than the numbers presented in Figure \ref{fig07}(a)), which provides values averaged over regions of the system.

\newpage
\section{Conclusions and remarks}


Fully atomistic reactive molecular dynamics (MD) simulations were carried out to investigate the dynamics of high-velocity ballistic impacts of BNNT-peapods against a solid target for the cases of vertical and horizontal collisions.
For low velocity vertical (1 km/s) shootings, the major structural deformations are on the BNNTs (mainly tube bending). Increasing the velocity the tube is fractured with C$_{60}$ ejections. For horizontal shootings, the stress is better distributed and BNNTs retain their structural integrity for small velocity values. Increasing the velocity values, the tube fracture became extensive, with bonds breaking mainly near the encapsulated fullerenes.
One interesting result is the formation of nanoribbon structures from unzipped tubes. For instance, for a velocity impact of 3.5 km/s, the BNNT unzips with the formation of a bilayer nanoribbon 'incrusted' with C$_{60}$.


\section{Acknowledgements}

This work was supported in part by the Brazilian Agencies CAPES, CNPq (process \#310045/2019-3) and FAPESP. J.M.S, M.M. C.F.W, L.D.M., P.A.S.A and D.S.G thank the Center for Computational Engineering and Sciences at Unicamp for financial support through the FAPESP/CEPID Grant \#2013/08293-7. JMS acknowledges CENAPAD-SP (Centro Nacional de Alto Desenpenho em São Paulo - Universidade Estadual de Campinas - UNICAMP) for the computational support process (proj842).

\bibliographystyle{elsarticle-num}
\bibliography{Peapods.bib}
\end{document}